\documentclass[preprint,eqsecnum,aps,nofootinbib]{revtex4}
\usepackage{amsfonts,amsmath,amssymb,amsthm}
\usepackage{latexsym}
\usepackage{bbm,bm}
\usepackage{graphicx}

\newcommand{\ket}[1]{\lvert #1 \rangle}
\newcommand{\bra}[1]{\langle #1 \lvert}
\newcommand{\beq}{\begin{equation}}
\newcommand{\eeq}{\end{equation}}
\newcommand{\beqs}{\begin{eqnarray}}
\newcommand{\eeqs}{\end{eqnarray}}
\begin{document}

\title{Average R\'{e}nyi Entropy of a Subsystem in Random Pure State}

\author{MuSeong Kim$^1$, Mi-Ra Hwang$^1$, Eylee Jung$^1$, and DaeKil Park$^{1,2}$\footnote{corresponding author, dkpark@kyungnam.ac.kr} }

\affiliation{$^1$Department of Electronic Engineering, Kyungnam University, Changwon,
                 631-701, Korea    \\
             $^2$Department of Physics, Kyungnam University, Changwon,
                  631-701, Korea }

%\author{DaeKil Park$^{1,2}$\footnote{dkpark@kyungnam.ac.kr} }
%
%\affiliation{$^1$Department of Electronic Engineering, Kyungnam University, Changwon
%                631-701, Korea    \\
%           $^2$Department of Physics, Kyungnam University, Changwon
%                 631-701, Korea    
%                      }

\begin{abstract}

In this paper we examine the average R\'{e}nyi entropy $S_{\alpha}$ of a subsystem $A$ when the whole composite system $AB$ is a 
random pure state. We assume that the Hilbert space dimensions of $A$ and $AB$ are $m$ and $m n$ respectively. 
First, we compute the average R\'{e}nyi entropy analytically for $m = \alpha = 2$. We compare this analytical result 
with the approximate average R\'{e}nyi entropy, which is shown to be very close. For general case we compute the average of the approximate R\'{e}nyi entropy $\widetilde{S}_{\alpha} (m,n)$ analytically. 
When $1 \ll n$,  $\widetilde{S}_{\alpha} (m,n)$ reduces to $\ln m - \frac{\alpha}{2 n} (m - m^{-1})$, which is in agreement with the asymptotic 
expression of the average von Neumann entropy. Based on the analytic result of  $\widetilde{S}_{\alpha} (m,n)$ we plot the $\ln m$-dependence of the quantum information derived from $\widetilde{S}_{\alpha} (m,n)$.
It is remarkable to note that the nearly vanishing region of the information becomes shorten with increasing $\alpha$, and eventually disappears in the limit of $\alpha \rightarrow \infty$. The physical implication of the result is briefly discussed.

\end{abstract}
\maketitle

\section{Introduction}
Although their motivations are different, the authors of Ref.\cite{lubkin,lloyd88,page93-1} considered a similar problem: the average von Neumann entropy of a subsystem $\rho_A$ whose Hilbert space dimension is $m$ when the whole system is a $m n$-dimensional random bipartite pure state 
$\rho = \ket{\psi}_{AB}\bra{\psi}$ with a condition $m \leq n$. Of course, $\rho_A = \mbox{Tr}_B \rho$ and $\rho_B = \mbox{Tr}_A \rho$. In particular, Ref.\cite{page93-1} introduced the probability distribution
\begin{equation}
\label{distribution-1}
P \left(p_1, \cdots, p_m \right) dp_1 \cdots dp_m \propto \delta \left( 1 - \sum_{i=1}^m p_i \right) \prod_{1 \leq i < j \leq m} (p_1 - p_j)^2 \prod_{k=1}^m \left( p_k^{n - m} dp_k \right)
\end{equation}
where $\{p_1, \cdots,p_m\}$ are eigenvalues of $\rho_A$. Thus, the problem can be summarized as a computation of the following quantity:
\begin{equation}
\label{avg-von-1}
S_{von} (m, n) \equiv \langle S_A \rangle = - \int \left( \sum_{i=1}^m p_i \ln p_i \right ) P \left(p_1, \cdots, p_m \right) dp_1 \cdots dp_m.
\end{equation}
Page in Ref.\cite{page93-1} computed $S_{von} (2,n)$ and $S_{von} (3,n)$ analytically, and  $S_{von} (4,n)$ and $S_{von} (5,n)$ with the aid of MATHEMATICA $2.0$.
Finally, he conjectured that $S_{von} (m,n)$ is 
\begin{equation}
\label{avg-von-2}
S_{von} (m, n) = \sum_{n = n + 1}^{mn} \frac{1}{k} - \frac{m-1}{2 n} \sim \ln m - \frac{m}{2 n}
\end{equation}
where the last equation is valid only for $1 \ll m \leq n$. The last term $\frac{m}{2 n}$ indicates that the entanglement entropy obeys a volume-law\cite{volume-law}. 
The Page's conjecture was rigorously proven in Ref.\cite{foong,jorge,sen96}. In particular, authors in Ref.\cite{jorge,sen96} changes the multiple integral of Eq. (\ref{avg-von-1}) into a single integral 
by using a generalized Laguerre polynomial\cite{table-1}.

In Ref.\cite{page93-2} Page applied Eq. (\ref{avg-von-2}) to the information loss problem\cite{hawk76,pre92} in the Hawking radiation\cite{hawk74,hawk75}.
He assumed that the whole random pure state $\ket{\psi}_{AB}$ represents the Hawking radiation ($\rho_A$) and the remaining black hole ($\rho_B$) states.
The reason why the random state is chosen is that the composite state is assumed to be highly complicate and hence, we do not know the state $\ket{\psi}_{AB}$ exactly.
Defining the quantum information $I_{von} (m,n) = \ln m - S_{von} (m,n)$, he plots the $\ln m$-dependence of $I_{von} (m,n)$ (see Fig. 2(b)) and claimed that the information may come out initially so slowly.
His calculation suggests that in order to obtain a sufficient information from Hawking radiation it takes at least the time necessary to radiate half the entropy of the black hole\cite{HP-1,YK-1}.
Research on this issue is not concluded and is still ongoing. 

The Page curve (\ref{avg-von-2}) is extended to the multipartite case\cite{ana-17,hwang-17} and random mixed states\cite{negativity-1}. Besides black hole, it is also applied to many different fields such as 
fermion systems\cite{balents-18,cirac22-1}, random spin chain\cite{paola22-1}, bosonic\cite{iosue-1} and fermion\cite{bianchi-21,nandy-21} Gaussian states, quantum thermalization\cite{yang15-1,vidmar-17,nakagawa-1,kaneko-19}, and 
quantum chaos\cite{vidmar-17,grover19,rigol20-1,shreya-1,sinha-21,telles-1}. It is also applied to the quantum information theories like random quantum circuits\cite{oliveria-07,douglas-08,bera20} and random quantum channels\cite{hayden08,horo10-1, fukuda10}.

In this paper we will extend Ref.\cite{page93-1} to the average R\'{e}nyi entropy defined as 
\begin{equation}
\label{avg-renyi-1}
S_{\alpha} (m, n) \equiv \langle S_{A,\alpha} \rangle = \frac{1}{1 - \alpha} \int \ln \left( \sum_{i=1}^m p_i^{\alpha} \right)  P \left(p_1, \cdots, p_m \right) dp_1 \cdots dp_m.
\end{equation}
Even though we apply the method of Ref.\cite{sen96}, it is impossible to convert the multiple integral of Eq. (\ref{avg-renyi-1}) into a single integral. 
Thus, it seems to be impossible to compute $S_{\alpha} (m,n)$ analytically. In the next section, however, we compute $S_{\alpha=2} (2,n)$ analytically. 
It was shown in this section that the analytical result of $S_{\alpha=2} (2,n)$ is very close to the approximate R\'{e}nyi entropy defined by
\begin{equation}
\label{avg-renyi-2}
\widetilde{S}_{\alpha} (m, n) = \frac{1}{1 - \alpha} \ln \left( \sum_{i=1}^m \langle  p_i^{\alpha} \rangle \right)
\end{equation}
when $m=\alpha = 2$. In Eq. (\ref{avg-renyi-2}) $\sum_{i=1}^m \langle  p_i^{\alpha} \rangle$ is defined as 
\begin{equation}
\label{avg-renyi-3}
 \sum_{i=1}^m \langle p_i^{\alpha} \rangle \equiv Z_{\alpha}  = \int \left( \sum_{i=1}^m p_i^{\alpha} \right)  P \left(p_1, \cdots, p_m \right) dp_1 \cdots dp_m.
\end{equation}
In section III we will compute $Z_{\alpha}$ explicitly for any positive real $\alpha$. It is represented as double summations. 
In section IV we compute the approximate R\'{e}nyi entropy $\widetilde{S}_{\alpha} (m, n)$ analytically. 
Defining the quantum information $I_{\alpha}(m,n) = \ln m - \widetilde{S}_{\alpha} (m, n)$ and using various asymptotic formula, we show that for large $n$
$I_{\alpha}(m,n)$ reduces to $\frac{\alpha}{2 n} (m - m^{-1})$, which is in agreement with Eq. (\ref{avg-von-2}) when $\alpha = 1$ and $m \gg 1$. We plot 
$I_{\alpha}(m,n)$ with varying $\alpha$ in this section and compare it to the case of von Neumann entropy presented in Ref. \cite{page93-2}. With increasing $\alpha$, the 
region for the almost vanishing information of $I_{\alpha} (m,n)$ becomes shorten and eventually disappears at $\alpha = \infty$. This means that in the application to the black hole radiation the quantum information of the R\'{e}nyi entropy comes out 
more earlier than that of von Neumann entropy. In section V a brief conclusion is given.

\section{computation of $S_{\alpha = 2}(2, n)$}

%%%%%%%%%%%%%%%%%%%%%%%%%%%%%%%%%%%%%%%%%%%%%%%%%%%%%%%%%
\begin{figure}[ht!]
\begin{center}
\includegraphics[height=10.0cm]{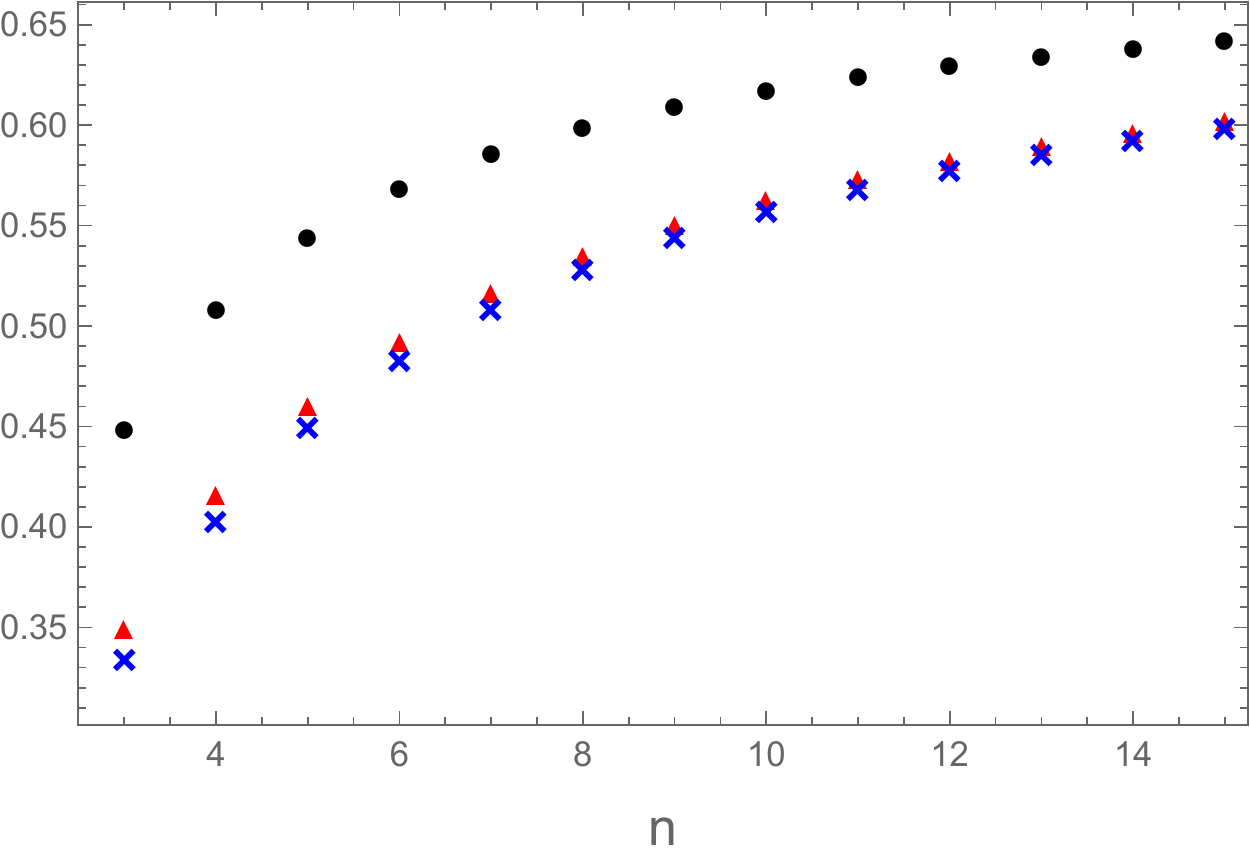} 

\caption[fig1]{(Color online) The $n$-dependence of exact R\'{e}nyi entropy (red triangle) given in Eq. (\ref{special-7}) and approximate 
 R\'{e}nyi entropy (blue cross) given in Eq. (\ref{special-8}) when $m = \alpha = 2$.  The black dot represents the exact von Neumann entropy given in Eq. (\ref{avg-von-2}) with $m=2$. }
\end{center}
\end{figure}
%%%%%%%%%%%%%%%%%%%%%%%%%%%%%%%%%%%%%%%%%%%%%%%%%%%%%%%%%%%

Defining $q_i = r p_i$, one can show from Eq. (\ref{avg-renyi-1}) that $S_{\alpha} (m,n)$ can be expressed as 
\begin{equation}
\label{special-1}
S_{\alpha}(m,n) = \frac{\alpha}{\alpha - 1} \psi (m n) - \frac{1}{\alpha - 1}
\frac{\int \ln \left( \sum_{i=1}^m q_i^{\alpha} \right) Q dq_1 \cdots dq_m}{\int  Q dq_1 \cdots dq_m}
\end{equation}
where $\psi (z) = \Gamma' (z) / \Gamma(z)$ is a digamma function and\footnote{Eq. (\ref{special-2}) is called a density of the eigenvalues of the Wishart matrix.}
\begin{equation}
\label{special-2}
Q (q_1, \cdots, q_m) dq_1 \cdots dq_m = \prod_{1 \leq i < j \leq m} (q_1 - q_j)^2 \prod_{k=1}^m \left(e^{-q_k} q_k^{n - m} dq_k \right).
\end{equation}

Now, we put $m = \alpha = 2$. In this case it is easy to show 
\begin{equation}
\label{special-3}
\int Q dq_1 dq_2 = \frac{2}{n -1} \Gamma^2 (n).
\end{equation}
Also the numerator in Eq. (\ref{special-1}) can be written as 
\begin{equation}
\label{special-4}
\int \ln (q_1^2 + q_2^2) Q dq_1 dq_2 = \int_0^{\infty} dq_1 \int_0^{\infty} dq_2 e^{-(q_1 + q_2)} \left[ q_1^n q_2^{n-2} + q_1^{n-2} q_2^n - 2 \left(q_1^{n-1} \right)^2 \right] \ln (q_1^2 + q_2^2).
\end{equation} 
In order to compute Eq. (\ref{special-4}) analytically, we use the following double integral formula\cite{table-2}
\begin{equation}
\label{double-int}
\int_0^{\infty} dx \int_0^{\infty} dy \ln (x^2 + y^2) e^{-p x - q y} = -\frac{2}{p q} \left[ \gamma + \frac{2 p^2 \ln q + 2 q^2 \ln p - \pi p q}{2 (p^2 + q^2)} \right]
\end{equation}
where $\gamma = 0.5772$ is Euler's constant.
Applying $\left( - \frac{\partial}{\partial p} \right)^m \left( - \frac{\partial}{\partial q} \right)^n$ to both sides of Eq. (\ref{double-int}) and putting $p=q=1$ at the final stage of calculation, one can compute
\begin{equation}
\label{special-5}
 F (m,n) = \int_0^{\infty} dx \int_0^{\infty} dy x^m y^n \ln (x^2 + y^2) e^{-x - y}
\end{equation}
analytically. For example, $F(2,3) = F(3, 2) = -24 \gamma + 21 \pi - 14$. In principle, the general expression of $F(m,n)$ for arbitrary integers $m$ and $n$ can be derived with the aid of 
MATHEMATICA $13.1$. Since, however, it is very lengthy and complicated\footnote{Furthernore, $F(m,n)$ depends on the $j^{th}$ term of some recurrence relations, where $j$ is a function of $m$ and $n$. This term is expressed with the aid of few 
special functions such as Lerch transcendent 
$$ \Phi (z,s,a) = \sum_{k=0}^{\infty} \frac{z^k}{(k + a)^s}.$$  }, we will not present the explicit expression in this paper. 

Using Eq. (\ref{special-5}) it is easy to show
\begin{equation}
\label{special-6}
\int \ln(q_1^2 + q_2^2) Q dq_1 dq_2 = 2 F(n, n-2) - 2 F(n-1, n-1).
\end{equation}
Inserting Eqs. (\ref{special-3}) and (\ref{special-6}) into Eq. (\ref{special-1}) with assuming $m = \alpha = 2$, the average R\'{e}nyi entropy becomes
\begin{equation}
\label{special-7}
S_{\alpha=2}(2,n) = 2 \psi (2 n) - (n - 1) \frac{F(n, n-2) - F(n-1, n-1)}{\Gamma^2 (n)}.
\end{equation}
As we will show later, one can show $\langle p_1^2 + p_2^2 \rangle = (n-2) / (2 n + 1)$. Therefore, Eq. (\ref{avg-renyi-2}) reduces to 
\begin{equation}
\label{special-8}
\widetilde{S}_{\alpha=2} (2, n) = -\ln \left(\frac{n + 2}{2 n + 1} \right).
\end{equation}

In Fig. 1 we plot the $n$-dependence of $S_{\alpha=2} (2,n)$ and $\widetilde{S}_{\alpha=2}(2,n)$ as red triangle and blue cross respectively. 
The black dot represent the average von Neumann entropy given in  Eq. (\ref{avg-von-2}) with $m=2$. As expected, the R\'{e}nyi entropy is less than the 
von Neumann entropy. As the figure shows, the exact and approximate R\'{e}nyi entropies are very close to each other.

\section{computation of $Z_{\alpha} = \sum_{j=1}^m \langle p_j^{\alpha} \rangle$}

In this section we will compute $\sum_{j=1}^m \langle p_j^{\alpha} \rangle$ analytically. First, we assume that $\alpha$ is integer for simplicity.
Later we will derive the expression of $Z_{\alpha}$ for any positive real $\alpha$. 
This will be used later to compute $\widetilde{S}_{\alpha} (m,n)$ presented in Eq. (\ref{avg-renyi-2}). 

Introducing $q_i = r p_i$ again, one can show
\begin{equation}
\label{avg-power-1}
\sum_{j=1}^m \langle p_j^{\alpha} \rangle = \frac{\Gamma(m n)}{\Gamma(m n + \alpha)} 
\frac{\int \left( \sum_{i=1}^{m}  q_i^{\alpha} \right) Q dq_1 \cdots dq_m} {\int Q dq_1 \cdots dq_m}.
\end{equation}

As Refs.\cite{jorge,sen96} shows, we first note
\begin{eqnarray}
\label{monde-1}
\prod_{1 \leq i < j \leq m} (q_1 - q_j)^2 = 
\left| \begin{array}{ccc}

p_{0}^{\beta} (q_{1})	& \cdots	& p_{0}^{\beta} (q_{m})    \\

p_{1}^{\beta} (q_{1})	& \cdots	& p_{1}^{\beta}(q_{m})     \\

\vdots		& \ddots	& \vdots                                           \\

p_{m-1}^{\beta} (q_{1})	& \cdots	& p_{m-1}^{\beta} (q_{m})

       \end{array}
\right|^2
\end{eqnarray}
where
\begin{equation}
\label{monde-2}
p_k^{\beta} (q) = \sum_{r=0}^k  \left( \begin{array}{c}  k \\  r  \end{array}   \right) (-1)^r
\frac{\Gamma(k + \beta + 1)}{\Gamma(k + \beta - r + 1)} q^{k - r} = (-1)^k k! L_k^{\beta} (q).
\end{equation}
In Eq. (\ref{monde-2}) $L_k^{\beta} (q)$ is a generalized Laguerre polynomial. 
It is worthwhile noting that Eq. (\ref{monde-1}) is valid for any real $\beta$. Thus, we can choose $\beta$ freely for convenience. 
Using the properties of the generalized Laguerre polynomial, one can show\cite{table-2,table-3} 
\begin{equation}
\label{integral-1}
\int_0^{\infty} dq e^{-q} q^{\beta} p_{k_1}^{\beta} (q) p_{k_2}^{\beta} (q) = \Gamma (k_1 + 1) \Gamma (k_1 + \beta +1) \delta_{k_1,k_2}
\end{equation}
and 
\begin{equation}
\label{integral-2}
\int_0^{\infty} dq e^{-q} q^{a-1} p_k^b (q) = (1 - a + b)_k \Gamma (a) (-1)^k
\end{equation}
where $(a)_k = a (a + 1) \cdots (a + k - 1)$. 

Now, let us define 
\begin{equation}
\label{def-1}
J_m \equiv \frac{\int \left( \sum_{i=1}^{m}  q_i^{\alpha} \right) Q dq_1 \cdots dq_m} {\int Q dq_1 \cdots dq_m}.
\end{equation}
First, we consider the case of $m=2$ for simplicity. In this case we choose $\beta = n - 2$. Using Eq. (\ref{monde-1}) and 
orthogonality condition (\ref{integral-1}) it is easy to show
\begin{equation}
\label{bunmo-1}
\int Q dq_1 dq_2 = 2! \left[ \int dq_1 e^{-q_1} q_1^{n-2} \left( p_0^{n-2} (q_1) \right)^2 \right] \left[ \int dq_2 e^{-q_2} q_2^{n-2} \left( p_1^{n-2} (q_2) \right)^2 \right].
\end{equation}
Similarly, it is straightforward to show
\begin{eqnarray}
\label{bunja-1}
&& \int \left(\sum_{i=1}^2 q_i^{\alpha} \right) Q dq_1 dq_2            \\   \nonumber
&& = 2! \Bigg[ \left\{ \int dq_1 e^{-q_1} q_1^{n-2} \left( p_0^{n-2} (q_1) \right)^2 \right\} \left\{ \int dq_2 e^{-q_2} q_2^{n + \alpha-2} \left( p_1^{n-2} (q_2) \right)^2 \right\}   \\   \nonumber
&&   \hspace{.8cm}      +             \left\{ \int dq_1 e^{-q_1} q_1^{n + \alpha-2} \left( p_0^{n-2} (q_1) \right)^2 \right\} \left\{ \int dq_2 e^{-q_2} q_2^{n-2} \left( p_1^{n-2} (q_2) \right)^2 \right\} \Bigg].
\end{eqnarray}
Inserting Eqs. (\ref{bunmo-1}) and (\ref{bunja-1}) into Eq. (\ref{def-1}) with $m=2$ and using Eqs. (\ref{integral-1}) and (\ref{integral-2}), one can show
\begin{equation}
\label{ratio-1}
J_2 = \sum_{k=0}^1 \frac{I_{k,\alpha} (n-2)} {\Gamma(k+1) \Gamma(k + n - 1)}
\end{equation}
where 
\begin{equation}
\label{ratio-1-boso}
I_{k,\alpha} (x) = \int d q e^{-q} q^{\alpha + x} \left( p_k^x (q) \right)^2.
\end{equation}
Therefore, the multiple integral in Eq. (\ref{def-1}) is changed into a single integral.

Now, we consider the general case. In this case we choose $\beta = n - m$. Similar calculation leads 
\begin{equation}
\label{ratio-2}
J_m = \sum_{k=0}^{m-1} \frac{I_{k, \alpha} (n - m)}{\Gamma (k+1) \Gamma (k+n - m + 1)}.
\end{equation}

Finally, we should compute $I_{k,\alpha} (x)$ analytically. First, we note the recursion relation $p_k^x (q) = p_k^{x+1} (q) + k p_{k-1}^{x+1} (q)$. 
Applying this recursion relation iteratively, one can derive
\begin{equation}
\label{recur-1}
p_k^x (q) = \sum_{i=0}^{\ell} \left( \begin{array}{c}  \ell  \\ i  \end{array} \right) (k - i + 1)_i p_{k - i}^{x + \ell} (q)
\end{equation}
for all nonnegative integer $\ell$. Choosing $\ell = \alpha$ and using the orthogonality condition (\ref{integral-1}), one can compute $I_{k,\alpha}(x)$, 
whose explicit expression is 
\begin{equation}
\label{deri-1}
I_{k,\alpha} (x) = \Gamma^2 (k+1) \sum_{i=0}^{\alpha} \left(\begin{array}{c}  \alpha \\ i \end{array} \right)^2 
\frac{\Gamma(k + \alpha + x - i + 1)}{\Gamma(k - i + 1)}.
\end{equation}
Thus, inserting Eq. (\ref{deri-1}) into Eq. (\ref{ratio-2}) one can derive $J_m$ as double summations. Then, Eq. (\ref{avg-power-1}) is expressed as
\begin{equation}
\label{final-11}
Z_{\alpha} \equiv \sum_{i=1}^m \langle p_i^{\alpha} \rangle = \frac{\Gamma(m n)}{\Gamma(m n + \alpha)} 
\sum_{k=0}^{m-1} \frac{\Gamma(k + 1)}{\Gamma(k + n - m + 1)} \sum_{i=0}^{\alpha} \left(\begin{array}{c}  \alpha \\ i \end{array} \right)^2 
\frac{\Gamma(k + n - m + \alpha - i + 1)}{\Gamma(k - i + 1)}.
\end{equation}

Eq. (\ref{final-11}) can be used to prove the Page's conjecture (\ref{avg-von-2}). From Eq. (\ref{final-11}) one can differentiate $Z_{\alpha}$ with respect to $\alpha$. 
Using $\Gamma'(z) = \Gamma(z) \psi(z)$ and $\psi(n) = -\gamma + \sum_{k=1}^{n-1} k^{-1}$, it is straightforward to show that 
$-\frac{\partial}{\partial \alpha} Z_{\alpha} |_{\alpha = 1}$ coincides with Eq. (\ref{avg-von-2}) exactly. 

Finally, let me derive a different expression of $Z_{\alpha}$, which is valid for any positive real $\alpha$. In the second summation of Eq. (\ref{final-11}) the actual upper bound 
of the parameter $i$ is $\min (\alpha,k)$ because if $k < \alpha$, $\Gamma(k-i+1)$ located in denominator diverges when $k+1 \leq i \leq \alpha$. In order to avoid this inconvenience, we introduce a 
new variable $j = k - i$, which runs from $-\alpha$ to $m-1$. In this case $\Gamma(k-i+1)$ is changed into $\Gamma(j+1)$, which goes to infinity for $j \leq -1$. 
In this reason  negative $j$ does not contribute to $Z_{\alpha}$. As a result, $Z_{\alpha}$ is expressed in the form:
\begin{equation}
\label{final-12}
Z_{\alpha} = \frac{\Gamma(m n) \Gamma^2 (\alpha + 1)}{\Gamma (m n + \alpha)} \sum_{k=0}^{m-1} \frac{\Gamma (k + 1)}{\Gamma (k + n - m + 1)}
\sum_{j=0}^{m-1} \frac{\Gamma(n - m + \alpha + 1 + j)}{\Gamma^2 (k - j + 1) \Gamma^2 (\alpha - k + j + 1) \Gamma(j + 1)}.
\end{equation}
Although this expression has similar problem when $j > k + 1$, $\alpha$ is not involved in the summation upper bound. Thus, Eq. (\ref{final-12}) is valid for any 
positive real $\alpha$. The exactly same expression was derived in Ref.\cite{bianchi-19-1}. However, Eq. (\ref{final-11}) is more convenient if $\alpha$ is integer and 
$\alpha \ll m$ because number of summation is very small compared to that of Eq. (\ref{final-12}).

\section{Quantum Information from $\widetilde{S}_{\alpha} (m,n)$}

%%%%%%%%%%%%%%%%%%%%%%%%%%%%%%%%%%%%%%%%%%%%%%%%%%%%%%%%%
\begin{figure}[ht!]
\begin{center}
\includegraphics[height=5.0cm]{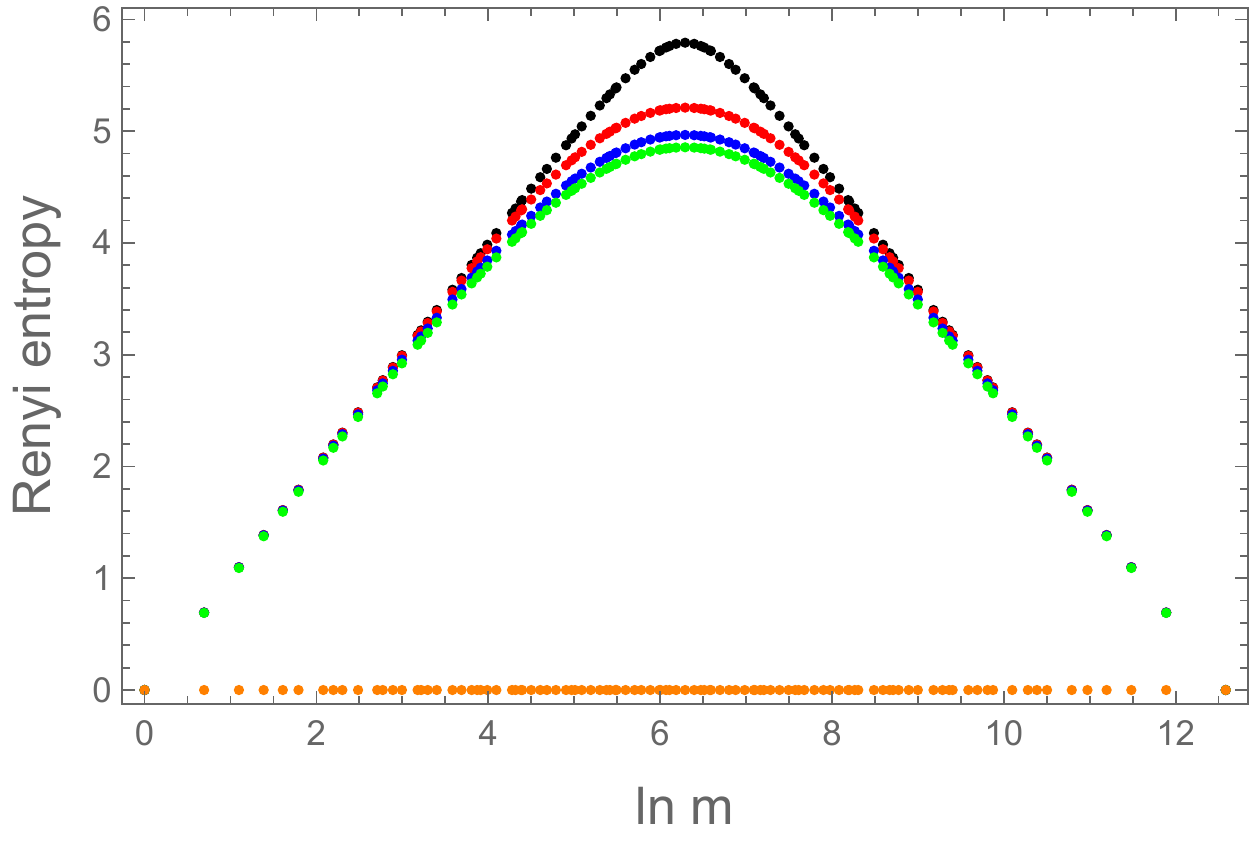} 
\includegraphics[height=5.0cm]{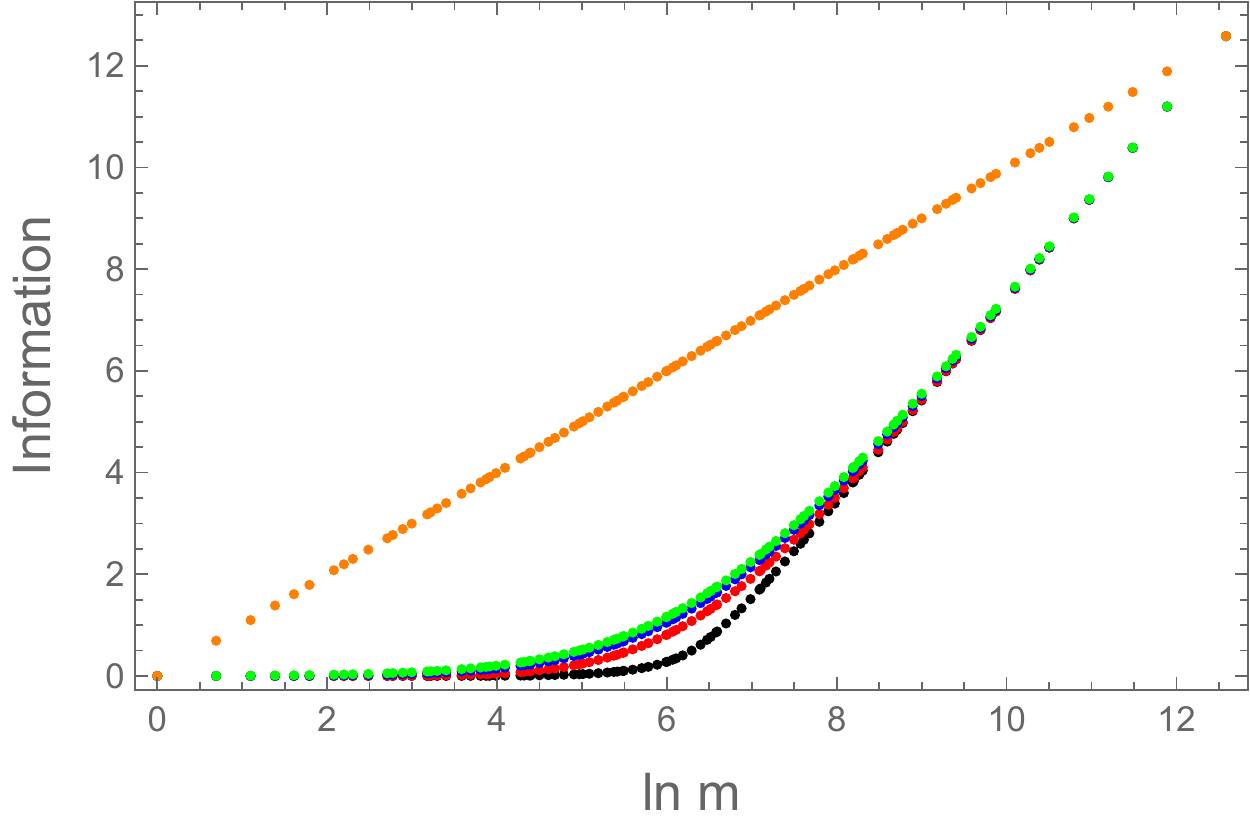}

\caption[fig2]{(Color online) $\ln m$-dependence of (a) $\widetilde{S}_{\alpha} (m,n)$ and (b) $I_{\alpha}(m,n)$. 
Here, we take $m n = 2^4 3^6 5^2 = 291600$. In both figures the black, red, blue, green, and orange dots correspond to $\alpha = 1$, $10$, $100$, $1000$, and $\infty$ respectively.  }
\end{center}
\end{figure}
%%%%%%%%%%%%%%%%%%%%%%%%%%%%%%%%%%%%%%%%%%%%%%%%%%%%%%%%%%%

From the previous sections the approximate R\'{e}nyi entropy $\widetilde{S}_{\alpha} (m,n)$ defined in Eq. (\ref{avg-renyi-2}) is given by
\begin{equation}
\label{info-1}
\widetilde{S}_{\alpha} (m,n) = \frac{1}{1 - \alpha} \ln Z_{\alpha}
\end{equation}
where $Z_{\alpha}$ is presented in Eq. (\ref{final-11}). Now, we assume $n \gg 1$. Using
\begin{eqnarray}
\label{info-2}
&&\lim_{z \rightarrow \infty} \Gamma(1 + z) \sim e^{-z} z^z \sqrt{2 \pi z} \left[ 1 + \frac{1}{12 z} + {\cal O} (z^{-2}) \right]   \\    \nonumber
&&\lim_{x \rightarrow \infty} \left(1 + \frac{a}{x} \right)^x \sim e^{a} \left[ 1 - \frac{a^2}{2 x} + {\cal O} (x^{-2}) \right],
\end{eqnarray}
one can show 
\begin{eqnarray}
\label{info-3}
&& \frac{\Gamma(m n)}{\Gamma(m n + \alpha)} \sim \left(m n \right)^{-\alpha)} \left[1 - \frac{\alpha (\alpha - 1)}{2 m n} + {\cal O} (n^{-2}) \right]   \\   \nonumber
&&\frac{\Gamma(n + k - m + 1 + \alpha - i)}{\Gamma (n + k - m + 1)}                                                                                                                                     \\   \nonumber
&& \hspace{1.0cm}  \sim n^{\alpha - i} \left[ 1 + \frac{1}{2 n} \left\{ i^2 + i (2 m - 2 k - 2 \alpha - 1) - \alpha (2 m - 2 k - \alpha - 1) \right\} + {\cal O} (n^{-2}) \right].
\end{eqnarray}
Then, for large $n$ $Z_{\alpha}$ reduces to 
\begin{equation}
\label{info-4}
Z_{\alpha} \sim m^{1 - \alpha} \left[ 1 + \frac{\alpha (\alpha - 1)}{2 n} (m - m^{-1}) + {\cal O} (n^{-2}) \right].
\end{equation}
As expected, $Z_{\alpha}$ becomes $m$ or $1$ when $\alpha = 0$ or $1$. 
%If $1 \ll m \ll n$, $Z_{\alpha}$ becomes $Z_{\alpha} \sim m^{1 - \alpha} \left[1 + \frac{m}{2 n} \alpha (\alpha - 1) +  {\cal O} (n^{-2}) \right]$.

Combining Eq. (\ref{info-1}) and Eq. (\ref{info-4}), for large $n$ $\widetilde{S}_{\alpha} (m,n)$ behaves as following:
\begin{equation}
\label{info-5}
\widetilde{S}_{\alpha} (m,n) \approx \ln m - \frac{1}{\alpha - 1} \ln \left[ 1 + \frac{\alpha (\alpha - 1)}{2 n} (m - m^{-1}) \right] \sim \ln m - \frac{\alpha}{2 n} (m - m^{-1}).
\end{equation}
If $1 \ll m$ and $\alpha = 1$, this equation reduces to $\widetilde{S}_{\alpha} (m,n) \sim \ln m - m / (2 n)$, which coincides with Eq. (\ref{avg-von-2}). 

The quantum information can be defined as the deficit of the average R\'{e}nyi entropy from the maximum:
\begin{equation}
\label{info-6}
I_{\alpha} (m,n) = \ln m - \widetilde{S}_{\alpha} (m,n) \sim \frac{1}{\alpha - 1} \ln \left[ 1 + \frac{\alpha (\alpha - 1)}{2 n} (m - m^{-1}) \right]
\end{equation}
where last equation is valid for $n \gg 1$.

Now, let us consider $m > n$ case. From the Schmidt decomposition we know that the eigenvalues of the density operators of systems A and B are the same. 
Thus, the approximate R\'{e}nyi entropy becomes $\widetilde{S}_{\alpha} (n,m)$. If $m \gg 1$, the information reduces to 
\begin{equation}
\label{info-7}
I_{\alpha} (m,n) \sim \ln m - \ln n + \frac{1}{\alpha - 1} \ln \left[ 1 + \frac{\alpha (\alpha - 1)}{2 m} (n - n^{-1}) \right].
\end{equation}

Finally, let us consider $\alpha \rightarrow \infty$ limit. Eq. (\ref{final-11}) implies that the leading term of $Z_{\alpha = \infty}$ is 
\begin{equation}
\label{inf_case}
Z_{\alpha = \infty} \sim \frac{\Gamma(m n)}{\Gamma(m) \Gamma(n)} \alpha^{-(m - 1) (n - 1)}. 
\end{equation}
Therefore, $\widetilde{S}_{\alpha \rightarrow \infty} = 0$ and $I_{\alpha = \infty} = \ln m$.

The $\ln m$-dependence of $\widetilde{S}_{\alpha} (m,n)$ and $I_{\alpha} (m,n)$ are plotted in Fig. 2 for $\alpha = 1$ (black), $10$ (red), $100$ (blue), $1000$ (green) and $\infty$ (orange). 
As Fig. 2(a) exhibits, the $\ln m$-dependence of the average R\'{e}nyi entropy $\widetilde{S}_{\alpha}(m,n)$ decreases with increasing $\alpha$, and eventually approaches to zero at $\alpha = \infty$. 
As Fig. 2(b) exhibits, the nearly vanishing region of $I_{\alpha} (m,n)$  is shorten with increasing $\alpha$ and eventually disappears at $\alpha = \infty$.

\begin{center}
\begin{tabular}{c|ccccc} \hline \hline
$\alpha \hspace{.5cm}$ &  $\hspace{.5cm} 1 \hspace{.5cm}$  & $\hspace{.5cm} 10 \hspace{.5cm}$  &  $\hspace{.5cm}100 \hspace{.5cm}$  &  $\hspace{.5cm}1000\hspace{.5cm}$  &  $ \hspace{.5cm} \infty$            \\   \hline  
$m_* \hspace{.5cm}$ & $243$   & $90$  &  $40$  &  $27$  &  $\hspace{.5cm} 2$                            \\      \hline  \hline
 \end{tabular}

\vspace{0.3cm}
Table I: The $\alpha$-dependence of $m_{*}$.
\end{center}
\vspace{0.5cm}

For example, let us define $m_{*}$, which is the smallest $m$ with satisfying $I_{\alpha} (m,n) > 0.1$. The $\alpha$-dependence of $m_*$ is summarized at Table I. 
As expected $m_*$ is decreasing with increasing $\alpha$.

\section{Conclusions}

In this paper we examine the average R\'{e}nyi entropy $S_{\alpha}$ of a subsystem $A$ when the whole composite system $AB$ is a 
random pure state. We assume that the Hilbert space dimensions of $A$ and $AB$ are $m$ and $m n$ respectively with $m \leq n$. 
If $m \geq n$, the Schmidt decomposition guarantees that the average value is obtained by simply interchanging $m$ and $n$.
First, we compute the average R\'{e}nyi entropy analytically for $m = \alpha = 2$. We compare this analytical result 
with the approximate average R\'{e}nyi entropy $\widetilde{S}_{\alpha = 2}(2,n)$. As Fig. 1 shows, these two results are very 
close to each other, especially when $n$ is large. For general case we compute $\widetilde{S}_{\alpha} (m,n)$ analytically. 
When $1 \ll n$,  $\widetilde{S}_{\alpha} (m,n)$ reduces to $\ln m - \frac{\alpha}{2 n} (m - m^{-1})$, which is in agreement with the asymptotic 
expression of the average von Neumann entropy given in Ref.\cite{page93-1}. Defining the information by Eq. (\ref{info-6}), we plot the $\ln m$ dependence of 
the information $I_{\alpha} (m,n)$ in Fig. 2(b). It is remarkable to note that the nearly vanishing region of $I_{\alpha} (m,n)$ becomes shorten with increasing $\alpha$, and eventually disappears 
in the limit of $\alpha \rightarrow \infty$. 

This result has important implication in the application of information loss problem. If we assume that $A$ and $B$ are the radiation and remaining black hole states, the information
derived from the R\'{e}nyi entropy can be obtained from Hawking radiation more and more earlier to that of von Neumann entropy with increasing $\alpha$, and
in the limit of $\alpha = \infty$ the information is radiated as soon as Hawking radiation starts. If this is right, we should re-consider the ``Alice and Bob'' thought-experiment described 
in Ref.\cite{gedanken,gedanken-2} on no-cloning theorem more carefully. Besides black hole physics, we want to examine the effect of our result in the quantum information theories like 
random quantum circuit and random quantum channel. 

The defect of our result is a fact that our calculation is based on $\widetilde{S}_{\alpha} (m,n)$. Although we guess $S_{\alpha} (m,n)$ also exhibits a similar behavior in Fig. 2, 
we can not prove it on the analytical ground. Numerical calculation is also very difficult when $m$ is large, because the calculation  requires $m$-multiple integration. 
Probably, we may need a new idea to explore this issue.

\vspace{1.0cm}

{\bf Acknowledgement}:
This work was supported by the National Research Foundation of Korea(NRF) grant funded by the Korea government(MSIT) (No. 2021R1A2C1094580).

%\vspace{1.0cm}

%{\bf Data Availability}:
% The datasets generated during and/or analyzed during the current study are available from the corresponding author on reasonable request.
 
% \vspace{1.0cm}

%{\bf Conflict of Interest}: 
%The authors declare that they have no known competing financial interests or personal relationships that could have appeared to influence the work reported in this paper.


\begin{thebibliography}{99}
\bibitem{lubkin} E. Lubkin, {\it Entropy of an $n$-system from its correlation with a $k$-reservoir}, J. Math. Phys. {\bf 19} (1978) 1028.
\bibitem{lloyd88} S. Lloyd and H. Pagels, {\it Complexity as Thermodynamic Depth}, Ann. Phy. {\bf 188} (1988) 186.
\bibitem{page93-1} D. N. Page, {\it Average Entropy of a Subsystem}, Phys. Rev. Lett. {\bf 71} (1993) 1291 [gr-qc/9305007].
\bibitem{volume-law} E. Bianchi, L. Hackl, M. Kieburg, M. Rigol, and L. Vidmar, {\it Volume-law entanglement entropy of typical pure quantum states}, PRX Quantum {\bf 3} (2022) 030201 [arXiv:2112.06959 (quant-ph)].
\bibitem{foong} S. K. Foong and S. Kanno, {\it Proof of Page's Conjecture on the Average Entropy of a Subsystem}, Phys. Rev. Lett. {\bf 72} (1994) 1148.
\bibitem{jorge} J. S\'{a}nchez-Ruiz, {\it Simple proof of Page's conjecture on the average entropy of a subsystem}, Phys. Rev. {\bf E 52} (1995) 5653.
\bibitem{sen96} S. Sen, {\it Average Entropy of a Subsystem}, Phys. Rev. Lett. {\bf 77} (1996) 1 [hep-th/9601132].
\bibitem{table-1} M. Abramowitz and I. A. Stegun, {\it Handbook of Mathematical Functions} (Dover Publications, New York, 1972).
\bibitem{page93-2} D. N. Page, {\it Information in Black Hole Radiation}, Phys. Rev. Lett. {\bf 71} (1993) 3743 [hep-th/9306083].
%\bibitem{hawking1} S. W. Hawking, {\it Breakdown of predictability in gravitational collapse}, Phys. Rev. {\bf D 14} (1976) 2460.
\bibitem{hawk76} S. W. Hawking, {\it Breakdown of Predictability in gravitational collapse}, Phys. Rev. {\bf D14} (1976) 2460.
\bibitem{pre92} J. Preskill, {\it Do black holes destroy information?} [hep-th/9209058].
\bibitem{hawk74} S. W. Hawking, {\it Black hole explosions?}, Nature {\bf 248} (1974) 30.
\bibitem{hawk75} S. W. Hawking, {\it Particle Creation by Black Holes}, Commun. Math. Phys. {\bf 43} (1975) 199.
\bibitem{HP-1} P. Hayden and J. Preskill, {\it Black holes as mirrors: quantum information in random subsystems}, J. High Energy Phys, {\bf 09} (2007) 120. [arXiv:0708.4025 (hep-th)].
\bibitem{YK-1}  B. Yoshida and A. Kitaev, {\it Efficient decoding for the Hayden-Preskill protocol}, arXiv:1710.03363 (hep-th).
\bibitem{ana-17} A. Alonso-Serrano and M. Visser, {\it Multipartite analysis of average-subsystem entropies}, Phys. Rev.{\bf A 96} (2017) 052302.
\bibitem{hwang-17}  J. Hwang, D. S. Lee, D. Nho, J. Oh, H. Park, D. Yeom, and H. Zoe, {\it Page curves for tripartite systems}, Class. Quant. Grav. {\bf 34} (2017) 145004 [arXiv:1608.03391 (hep-th)].
\bibitem{negativity-1} H. Shapourian, S. Liu, J. Kudler-Flam, and A. Vishwanath, {\it Entanglement negativity spectrum of random mixed states: A diagrammatic approach}, PRX Quantum {\bf 2} (2021) 030347 [arXiv:2011.01277 (cond-mat)].

\bibitem{balents-18} C. Liu, X. Chen, and L. Balents, {\it Quantum Entanglement of the Sachdev-Ye-Kitaev Models}, Phys. Rev. {\bf B 97} (2018) 245126 [arXiv:1709.06259 (cond-mat)].
\bibitem{cirac22-1} X. Yu, Z. Gong, and J. I. Cirac, {\it Free-fermion Page Curve: Canonical Typicality and Dynamical Emergence}, arXiv:2209.08871 (quant-ph).

\bibitem{paola22-1}P. Ruggiero and X. Turkeshi, {\it Quantum information spreading in random spin chains},  Phys. Rev. {\bf B 106} (2022) 134205 [arXiv:2206.02934 (cond-mat)].

\bibitem{iosue-1} J. T. Iosue, A. Ehrenberg, D. Hangleiter, A. Deshpande, and A. V. Gorshkov, {\it Page curves and typical entanglement in linear optics}, arXiv:2209.06838 (quant-ph).

\bibitem{bianchi-21} E. Bianchi, L. Hackl, and M. Kieburg, {\it The Page Curve for Fermionic Gaussian States}, Phys. Rev. {\bf B 103} (2021) 241118 [arXiv:2103.05416 (quant-ph)].
\bibitem{nandy-21} B. Bhattacharjee, P. Nandy,  and T. Pathak, {\it Eigenstate capacity and Page curve in fermionic Gaussian states}, Phys. Rev. {\bf B 104} (2021) 214306 [arXiv:2109.00557 (quant-ph)].

\bibitem{yang15-1} Z. Yang, C. Chamon, A. Hamma, and E. R. Mucciolo, {\it Two-component Structure in the Entanglement Spectrum of Highly Excited States}, Phys. Rev. Lett. {\bf 115} (2015) 267206 [arXiv:1506.01714 (cond-mat)].
\bibitem{vidmar-17} L. Vidmar and M. Rigol, {\it Entanglement Entropy of Eigenstates of Quantum Chaotic Hamiltonians}, Phys. Rev. Lett. {\bf 119} (2017) 220603 [arXiv:1708.08453  (cond-mat)].
\bibitem{nakagawa-1} Y. O. Nakagawa, M. Watanabe, H. Fujita, and S. Sugiura, {\it Universality in volume-law entanglement of scrambled pure quantum states}, Nat. Commun. {\bf 9} (2018) 1635.
\bibitem{kaneko-19} K. Kaneko, E. Iyoda, and T. Sagawa, {\it Characterizing complexity of many-body quantum dynamics by higher-order eigenstate thermalization}, Phys. Rev. {\bf A 101} (2020) 042126 [arXiv:1911.10755 (cond-mat)].



\bibitem{grover19} T. Lu and T. Grover, {\it Renyi Entropy of Chaotic Eigenstates}, Phys. Rev. {\bf E 99} (2019) 032111 [arXiv:1709.08784 (cond-mat)].
\bibitem{rigol20-1} P. \L yd\.{z}ba, M. Rigol, and L. Vidmar, {\it Eigenstate Entanglement Entropy in Random Quadratic Hamiltonians}, Phys. Rev. Lett. {\bf 125} (2020) 180604 [arXiv:2006.11302 (cond-mat)].
\bibitem{shreya-1} H. Liu and S. Vardhan, {\it A dynamical mechanism for the Page curve from quantum chaos}, J. High Energy Phys. {\bf 2021} (2021) 88 [arXiv:2002.05734 (hep-th)].
\bibitem{sinha-21} S. Sinha, S. Ray and S. Sinha, {\it Fingerprint of chaos and quantum scars in kicked Dicke model: An out-of-time-order correlator study}, arXiv:2101.05155 (cond-mat).
\bibitem{telles-1} J. T. de Miranda and T. Micklitz, {\it Subsystem Trace-Distances of Random States}, arXiv:2210.03213 (quant-ph).

\bibitem{oliveria-07} R. Oliveira, O. C. O. Dahlsten, and M. B. Plenio, {\it Generic Entanglement Can Be Generated Efficiently}, Phys. Rev. Lett. {\bf 98} (2007) 130502 [quant-ph/0605126].
\bibitem{douglas-08} A. D. K. Plato, O. C. Dahlsten, and M. B. Plenio, {\it Random circuits by measurements on weighted graph states}, Phys. Rev. {\bf A 78} (2008) 042332 [arXiv:0806.3058 (quant-ph)].
\bibitem{bera20} A. Bera and S. S. Roy, {\it Growth of genuine multipartite entanglement in random unitary circuits}, Phys. Rev. {\bf A 102} (2020) 062431 [arXiv:2003.12546 (quant-ph)].

\bibitem{hayden08} P. Hayden and A. Winter, {\it Counterexamples to the maximal $p$-norm multiplicativity conjecture for all $p > 1$}, Comm. Math. Phys. {\bf 284} (2008) 263 [arXiv:0807.4753 (quant-ph)].
\bibitem{horo10-1} F. G. S. L. Brandao and M. Horodecki, {\it On Hastings' counterexamples to the minimum output entropy additivity conjecture}, Open Syst. Inf. Dyn. {\bf 17} (2010) 31 [arXiv:0907.3210 (quant-ph)].
\bibitem{fukuda10} M. Fukuda and C. King, {\it Entanglement of random subspaces via the Hastings bound}, J. Math. Phys. {\bf 51} (2010) 042201 [arXiv:0907.5446 (quant-ph)].


\bibitem{table-2} A. P. Prudnikov, Y. A. Brychkov, and O. I. Marichev, {\it Integrals and Series} (Gordon and Breach Science Publishers, New York, 1986).

\bibitem{table-3} I. S. Gradshteyn and I. M. Ryzbik, {\it Table of Integrals, Series, and Products} (Academic Press, San Diego, 2000).

\bibitem{bianchi-19-1} E. Bianchi and P. Don\`{a}, {\it Typical entanglement entropy in the presence of a center: Page curve and its variance}, Phys. Rev. {\bf D 100} (2019) 105010 [arXiv:1904.08370 (hep-th)].
\bibitem{gedanken} L. Susskind and L. Thorlacius, {\it Gedanken Experiments involving Black Holes}, Phys. Rev. {\bf D49} (1994) 966 [hep-th/9308100]

\bibitem{gedanken-2} Y. Sekino and L. Susskind, {\it Fast Scramblers}, J. High Energy Phys. {\bf 0810} (2008) 065 [arXiv:0808.2096 (hep-th)].



















\end{thebibliography}
\end{document}